\documentclass[aps,prb,twocolumn,groupedaddress]{revtex4}

\usepackage{graphicx}
\usepackage{dcolumn}
\usepackage{bm}
\usepackage{subfigure}

\begin{document}

%\preprint{}

\title{Finite-Size Scaling in the Energy-Entropy Plane for the 2D
$\pm J$ Ising Spin Glass}

% address or url should go in the {}'s for \email and \homepage.4

\author{Ronald Fisch}
\email[]{ron@princeton.edu}
%\thanks{}
\affiliation{382 Willowbrook Dr.\\
North Brunswick, NJ 08902}


\date{\today}

\begin{abstract}
For $L \times L$ square lattices with $L \le 20$ the 2D Ising spin
glass with +1 and -1 bonds is found to have a strong correlation
between the energy and the entropy of its ground states.  A fit to
the data gives the result that each additional broken bond in the
ground state of a particular sample of random bonds increases the
ground state degeneracy by approximately a factor of 10/3.  For $x
= 0.5$ (where $x$ is the fraction of negative bonds), over this
range of $L$, the characteristic entropy defined by the
energy-entropy correlation scales with size as $L^{1.78(2)}$.
Anomalous scaling is not found for the characteristic energy,
which essentially scales as $L^2$.  When $x= 0.25$, a crossover to
$L^2$ scaling of the entropy is seen near $L = 12$.  The results
found here suggest a natural mechanism for the unusual behavior of
the low temperature specific heat of this model, and illustrate
the dangers of extrapolating from small $L$.

\end{abstract}

\pacs{75.10.Nr, 75.40.Mg, 75.50.Lk}
%\keywords{finite-size scaling, Ising spin glass}

\maketitle

\section{INTRODUCTION}

The Edwards-Anderson (EA) spin glass\cite{EA75} has been studied
extensively for thirty years.  A complete understanding of its
behavior in two and three dimensions remains elusive.  In recent
years it has become possible to compute the free energy of the
two-dimensional (2D) Ising spin glass with $\pm J$ bonds on $L
\times L$ lattices with $L$ of 100 or
more.\cite{BP91,BGP98,Hou01,HY01,LGMMR04,CHK04}  From these
calculations on large lattices we have learned that extrapolations
of data from lattices with $L < 30$ are often
misleading.\cite{SK93,SK94,SM96,MSS98,SMS00}

A better understanding of why this happens is clearly desirable.
This is especially true because essentially all of the work on
three-dimensional (3D) EA models at low temperatures must be done
on lattices with $L \le 20$, due to our inability to equilibrate
larger lattices at low temperatures in 3D.\cite{HG02}  At least
one example of complex behavior of the order parameter emerging as
$L$ is increased is already known in a similar 3D
model.\cite{Fis95}

In this work we will analyze data for the energies and entropies
of the ground states (GS) of 2D Ising spin glasses obtained using
methods from earlier work.\cite{SK93,SK94,LC01}  We will
demonstrate that for small square lattices the $\pm J$ EA model
has a strong correlation of the sample-to-sample fluctuations of
the energy and the entropy of the GS. The increase of GS entropy
$S_0$ with GS energy $E_0$ is too large to be explained by
fluctuations in the number of zero-energy single-spin flips.  This
correlation may be the cause of the breakdown of naive scaling
behavior at small $L$ in this model.

\section{THE MODEL}

The Hamiltonian of the EA model for Ising spins is
\begin{equation}
  H = - \sum_{\langle ij \rangle} J_{ij} {S}_{i} {S}_{j}   \, ,
\end{equation}
where each spin ${S}_{i}$ is a dynamical variable which has two
allowed states, +1 and -1.  The $\langle ij \rangle$ indicates a
sum over nearest neighbors on a simple square lattice of size $L
\times L$.  We choose each bond $J_{ij}$ to be an independent
identically distributed quenched random variable, with the
probability distribution
\begin{equation}
  P ( J_{ij} ) = x \delta (J_{ij} + 1)~+~(1 - x) \delta (J_{ij} -
  1)   \, ,
\end{equation}
so that we actually set $J = 1$, as usual.

The data analyzed here used an ensemble in which, for a given
value of $x$, every $L \times L$ random lattice sample had exactly
$(1 - x) L^2$ positive bonds and $x L^2$ negative bonds. Details
of the methods used to calculate $E_0$ and the numbers of GS have
been described earlier.\cite{LC01}  $S_0$ is defined as the
natural logarithm of the number of ground states. For each sample,
$E_0$ and $S_0$ were calculated for the four combinations of
periodic (P) and antiperiodic (A) toroidal boundary conditions
along each of the two axes of the square lattice.\cite{LC01}  We
will refer to these as PP, PA, AP and AA.  We use ALL to refer to
a data set which includes the results from all four types of
boundary conditions.  In the spin-glass region of the phase
diagram, the variation of the sample properties for changes of the
boundary conditions is small compared to the variation between
different samples of the same size,\cite{SK94} except when $x$ is
close to the ferromagnetic phase boundary and the ferromagnetic
correlation length becomes comparable to $L$.

\section{GROUND STATE PROPERTIES}

\begin{figure*}
\centering
\includegraphics[width=3in]{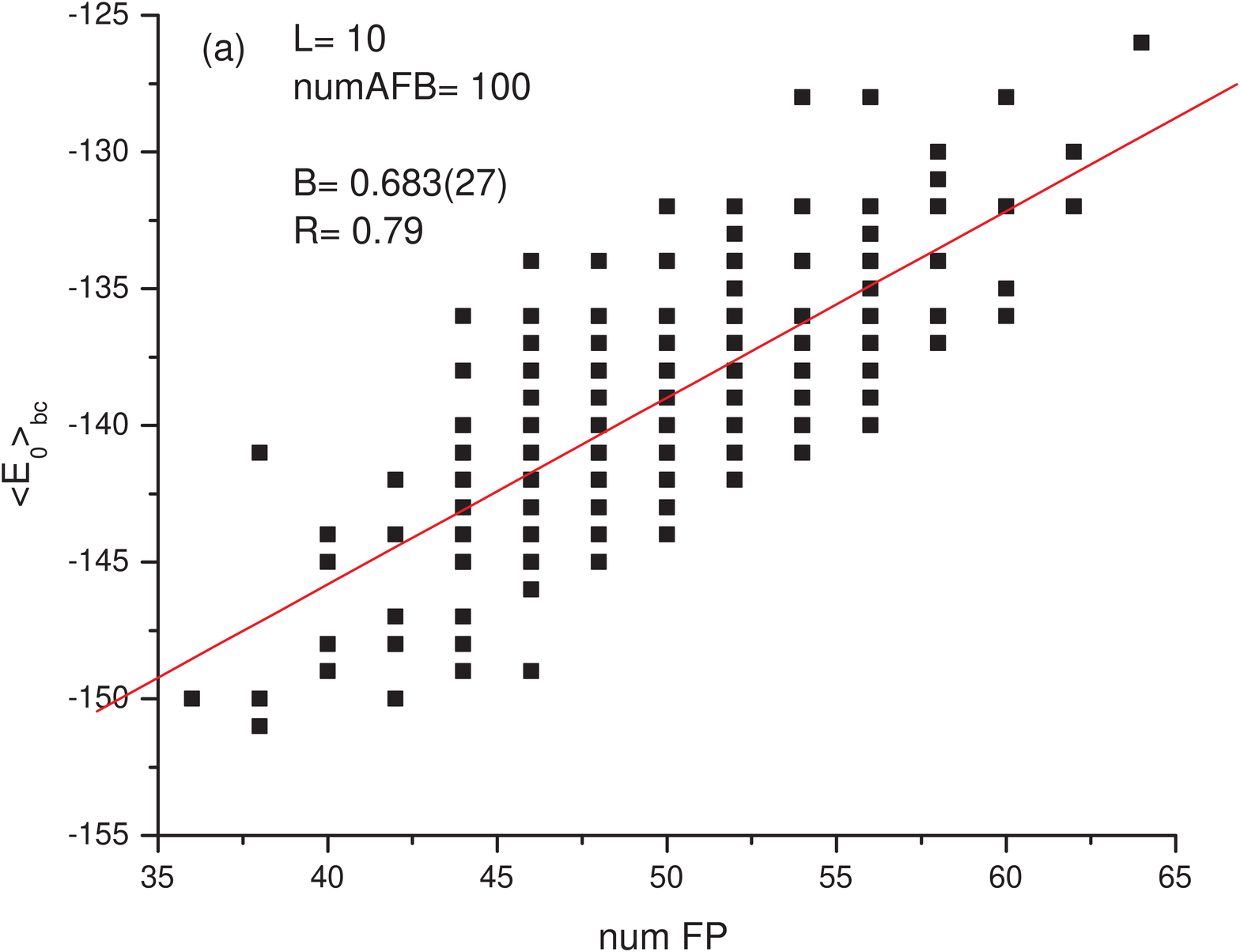}\quad
\includegraphics[width=3in]{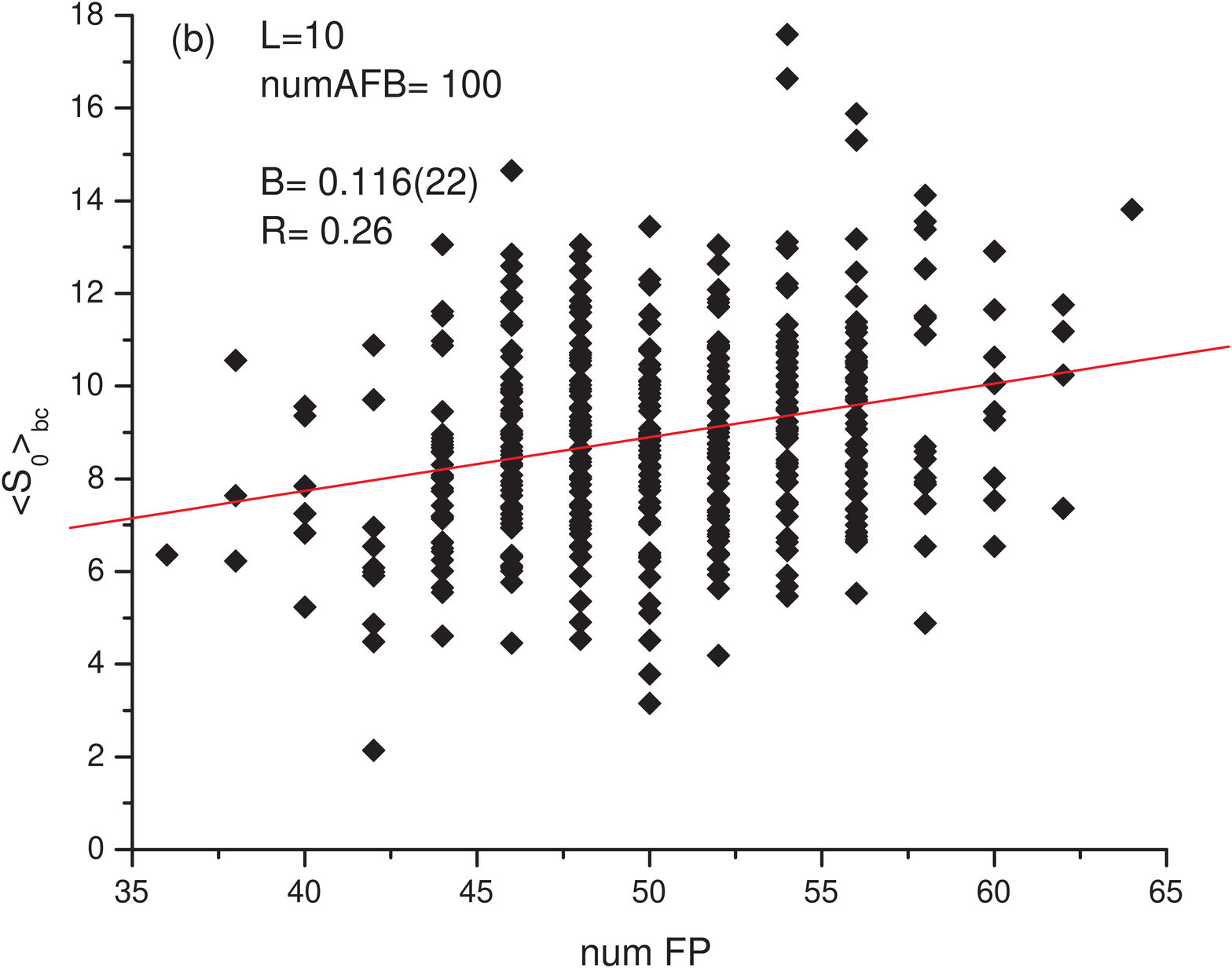}\quad
\includegraphics[width=3in]{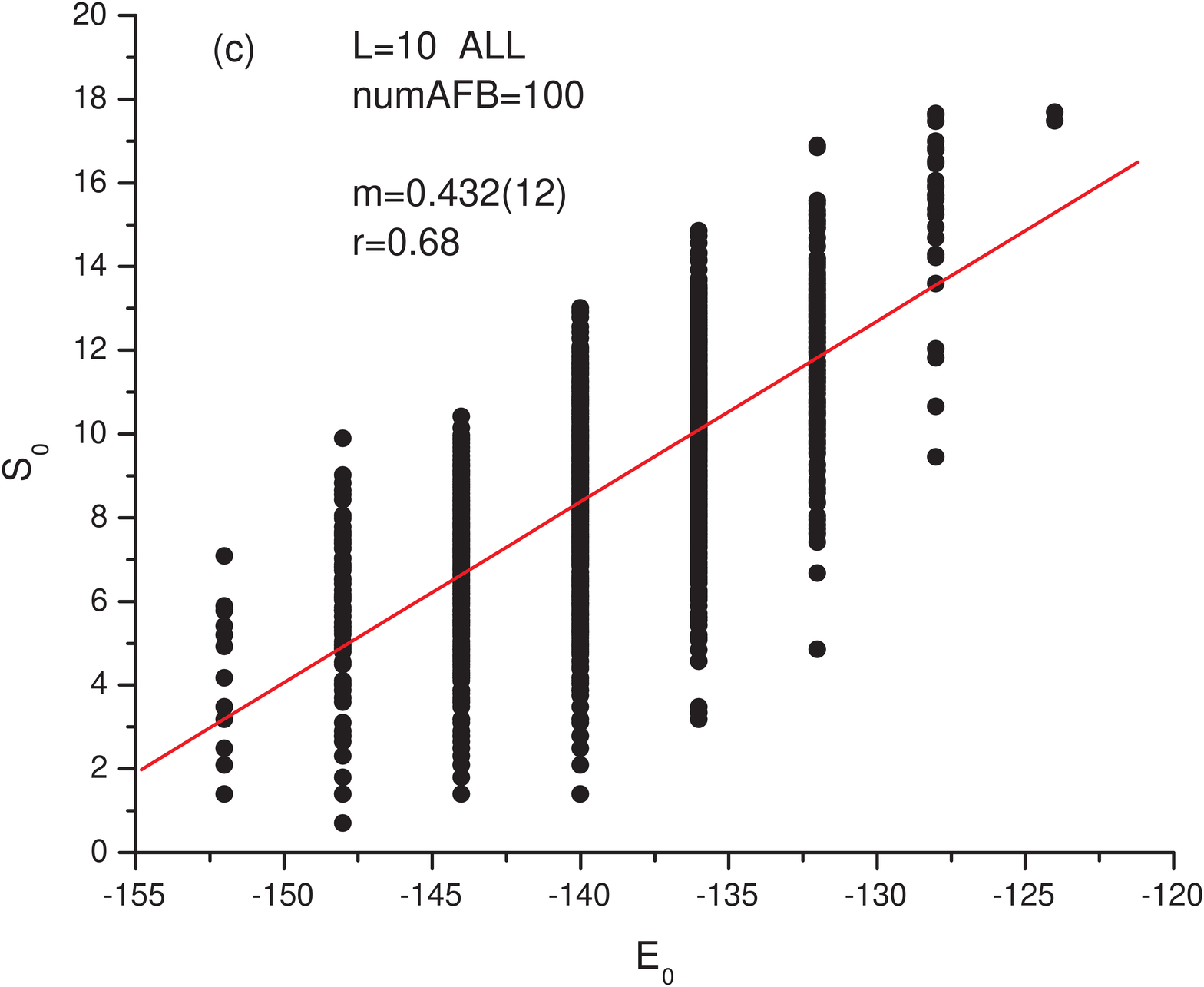}\quad
\includegraphics[width=3in]{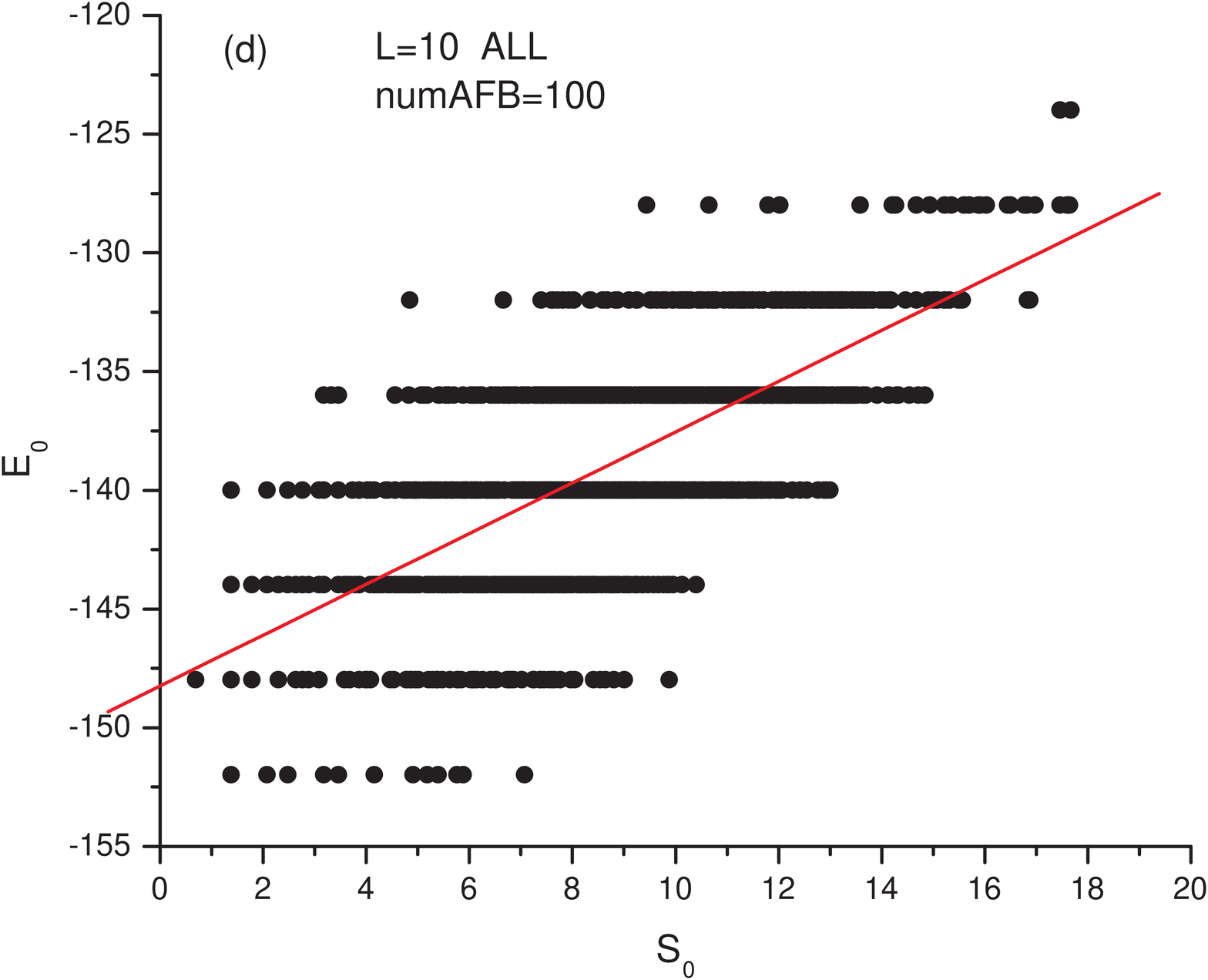}
\caption{\label{Fig.1}(color online) Scatter plots of correlations
for $x = 0.5$ and $L = 10$: (a) $E_0$ averaged over boundary
conditions vs. number of frustrated plaquettes; (b) $S_0$ averaged
over boundary conditions vs. number of frustrated plaquettes; (c)
$S_0$ vs. $E_0$; (d) $E_0$ vs. $S_0$. The number of samples used
is 400, and the lines through the data are least-squares fits.}
\end{figure*}

The average GS entropy $\langle S_0 \rangle$ of an $L \times L$
sample for this model is essentially proportional to $L^2$, the
number of spins, with a small finite-size correction.\cite{LC01}
It was discovered earlier,\cite{LC01} however, that for $x = 0.5$
the ratio of the width of the distribution of $S_0$ for different
samples of size $L$ divided by $\langle S_0 \rangle$ is not a
monotonic function of $L$, having a peak at $L = 8$.  A similar
change in behavior between $L=8$ and $L=10$ was seen earlier by
Saul and Kardar in samples with open boundary conditions, and
appears in Fig.~11 of their paper.\cite{SK94}  The original
motivation of the current study was to understand the origin of
this unexpected behavior.  Our ensemble, unlike the one used by
Saul and Kardar, does not keep the number of frustrated plaquettes
fixed.

We first look to see if the GS properties are correlated with the
number of frustrated plaquettes, with the number of bonds of each
type held fixed.  The scatter-plot data for $x = 0.5$ and $L = 10$
are shown in Fig.~1(a).  There is a substantial correlation of
$E_0$ with the number of frustrated plaquettes, and this
correlation seems to be independent of $L$.  Since it is well
known that $E_0$ increases as the number of frustrated plaquettes
is increased, this is expected.

There is a weaker correlation between $S_0$ and the number of
frustrated plaquettes, as shown in Fig.~1(b).  On the average,
increasing the number of frustrated plaquettes increases $S_0$.
This correlation is also not surprising, since positive $S_0$
arises from rearranging the strings of broken bonds which connect
the frustrated plaquettes in a GS.  It seems natural that a larger
number of frustrated plaquettes would give a larger number of ways
to rearrange the strings of broken bonds.

For Fig.~1(a) and 1(b), we have averaged $E_0$ and $S_0$ over the
four different boundary conditions for each sample, because the
number of frustrated plaquettes does not depend on the boundary
conditions. In the remainder of this work, we will treat each
boundary condition for each sample independently.

All equilibrium statistical mechanics can be derived from the
partition function, which is determined by the energy and the
entropy.  Therefore, we would like to know if $E_0$ and $S_0$ are
correlated with each other. The scatter plots for this correlation
from the same data are shown in Fig.~1(c), along with a
least-squares fit to the data, treating $E_0$ as the independent
variable.  Fig.~1(d) shows the same data with the roles of energy
and entropy reversed.  It demonstrates that the least-squares fit
depends on which variable is chosen as the independent one.

\begin{figure}
\includegraphics[width=3in]{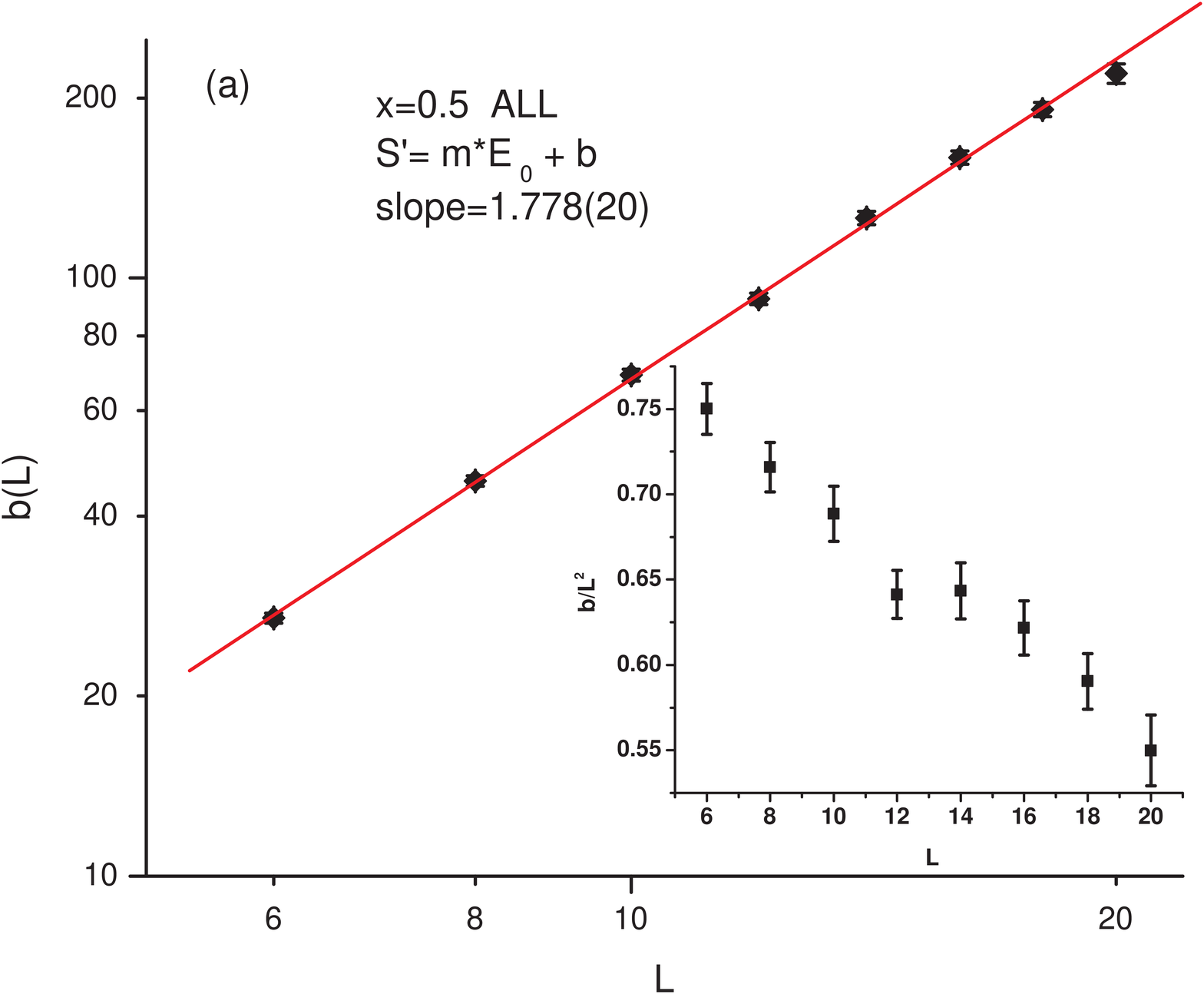}\quad
\includegraphics[width=3in]{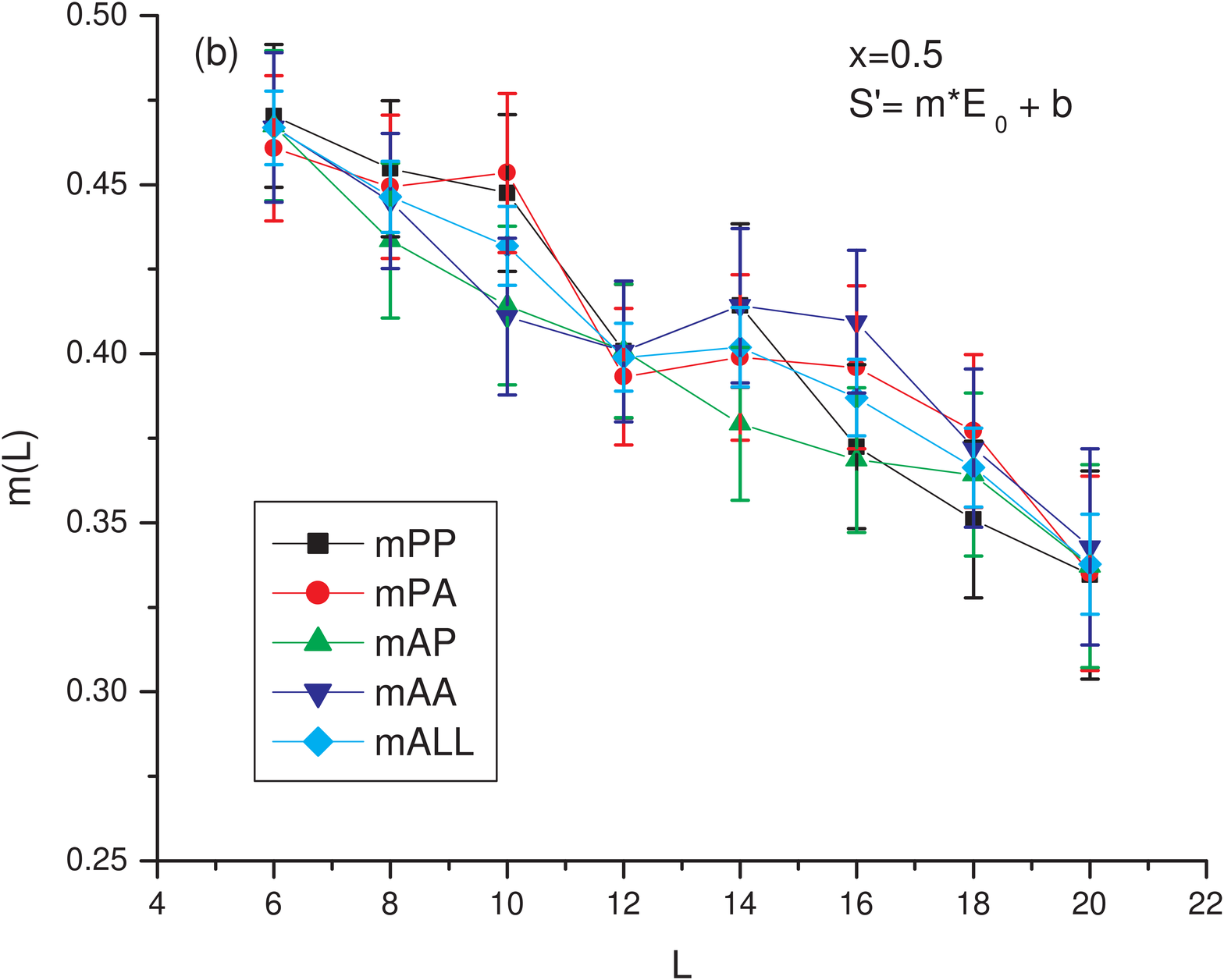}
\caption{\label{Fig.2}(color online) Results of least-squares
fit analysis parameterized by Eqn.~(3) of the scatter plot of
correlations between $E_0$ and $S_0$ for $x = 0.5$: (a)$b$ vs.
$L$, log-log plot; inset: $b / L^2$ vs. $L$; (b)$m$ vs. $L$.
The number of samples used for each $L$, $(L, \#)$, is (6:400),
(8:400), (10:400), (12:400), (14:400), (16:400), (18:400) and
(20:238).}
\end{figure}

\begin{figure}
\includegraphics[width=3in]{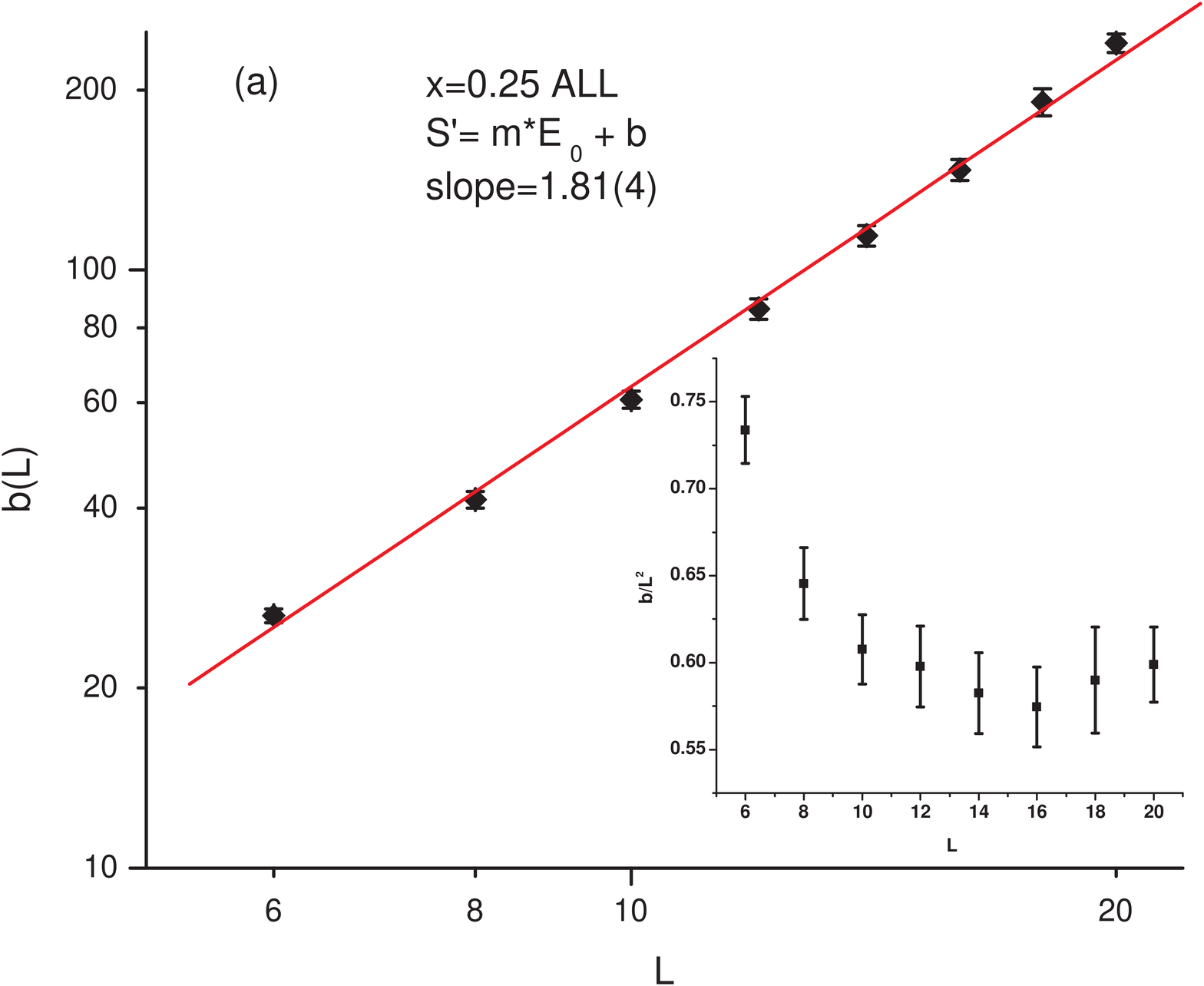}\quad
\includegraphics[width=3in]{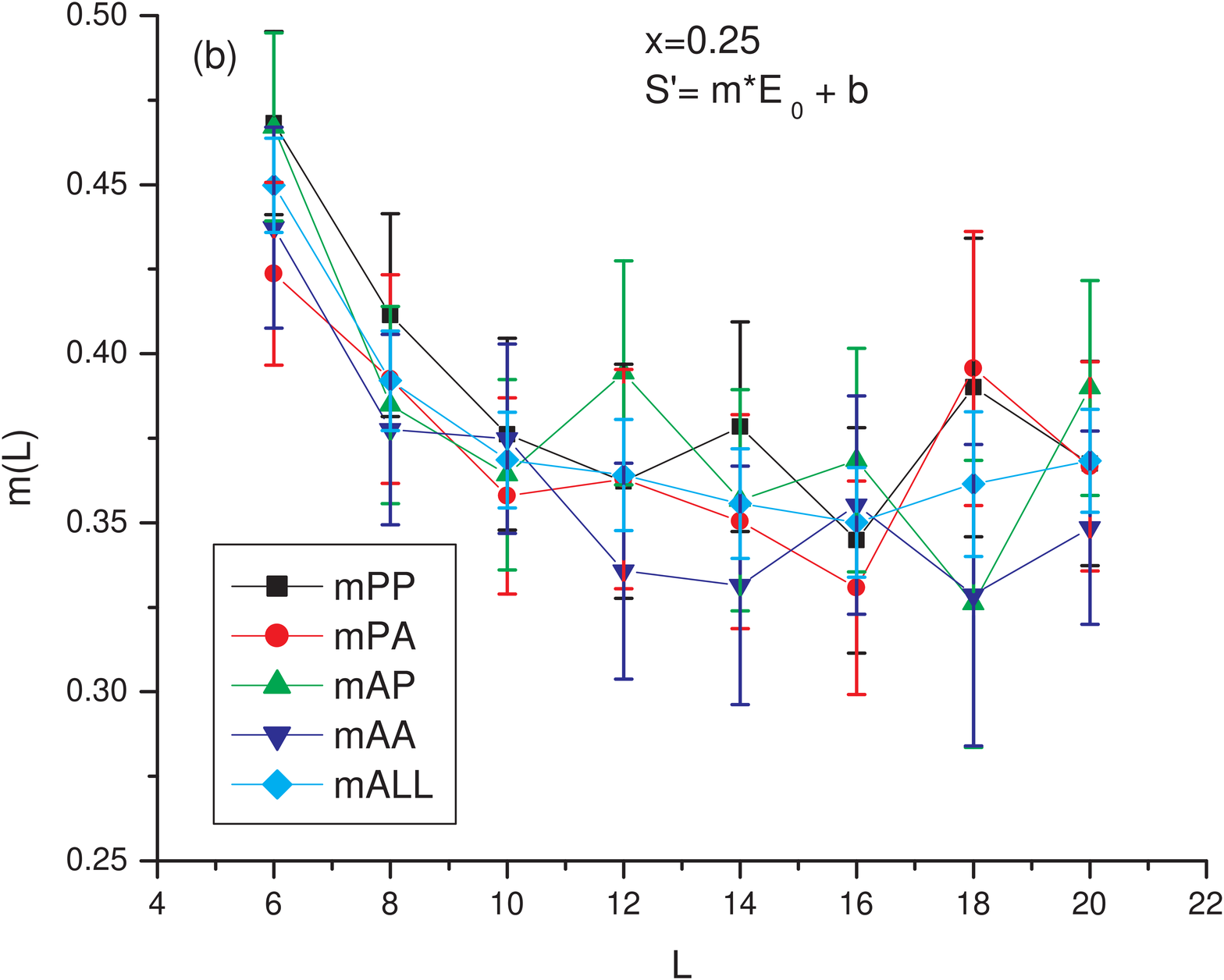}
\caption{\label{Fig.3}(color online) Results of least-squares
fit analysis parameterized by Eqn.~(3) of the scatter plot of
correlations between $E_0$ and $S_0$ for $x = 0.25$: (a)$b$ vs.
$L$, log-log plot; inset: $b / L^2$ vs. $L$; (b)$m$ vs. $L$.  The
number of samples used for each $L$, $(L, \#)$, is (6:200),
(8:200), (10:200), (12:200), (14:200), (16:200), (18:133) and
(20:200).}
\end{figure}

\begin{figure}
\includegraphics[width=3in]{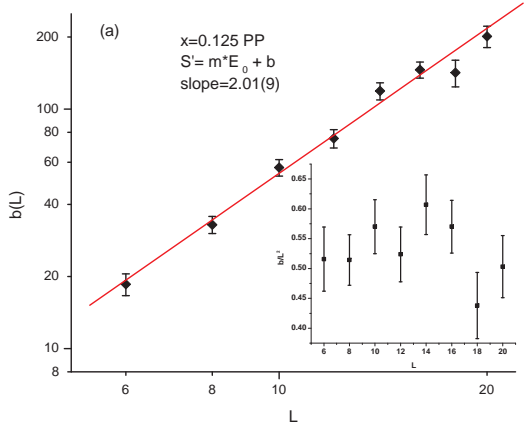}\quad
\includegraphics[width=3in]{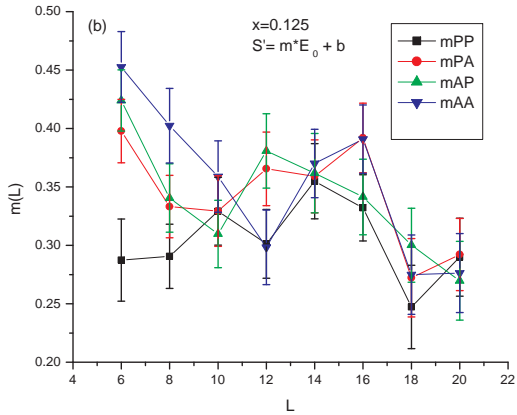}
\caption{\label{Fig.4}(color online) Results of least-squares
fit analysis parameterized by Eqn.~(3) of the scatter plot of
correlations between $E_0$ and $S_0$ for $x = 0.125$: (a)$b$ vs.
$L$, log-log plot; inset: $b / L^2$ vs. $L$; (b)$m$ vs. $L$.
The number of samples used for each $L$, $(L, \#)$, is (6:200),
(8:200), (10:200), (12:200), (14:200), (16:200), (18:200) and
(20:200).}
\end{figure}

The results of least-squares fits for $x = 0.5$, 0.25 and 0.125,
and $L$ varying from 6 to 20 are shown in Figures 2, 3 and 4,
respectively. For each value of $x$ and each $L$, we show the
slope $m$ given by the least-squares fit, and the offset $b$ of
the entropy, defined by
\begin{equation}
  S' (E_0) = m*E_0 + b  \, .
\end{equation}
$S'$ is the dependent variable in the least-squares fit.  Note
that $E_0$ is negative.

\begin{figure}
\includegraphics[width=3in]{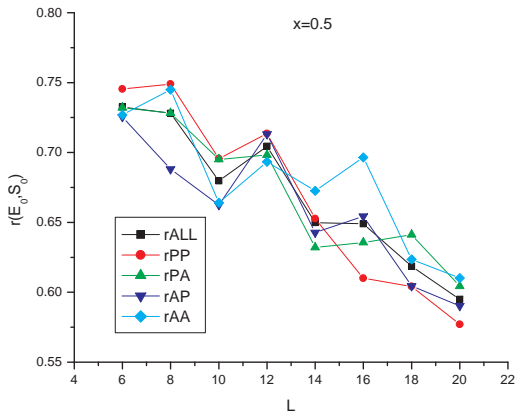}
\caption{\label{Fig.5}(color online) Covariance of ground state
energy and ground state entropy, $r(E_0,S_0)$ vs. $L$ for $x = 0.5$.}
\end{figure}

\begin{figure}
\includegraphics[width=3in]{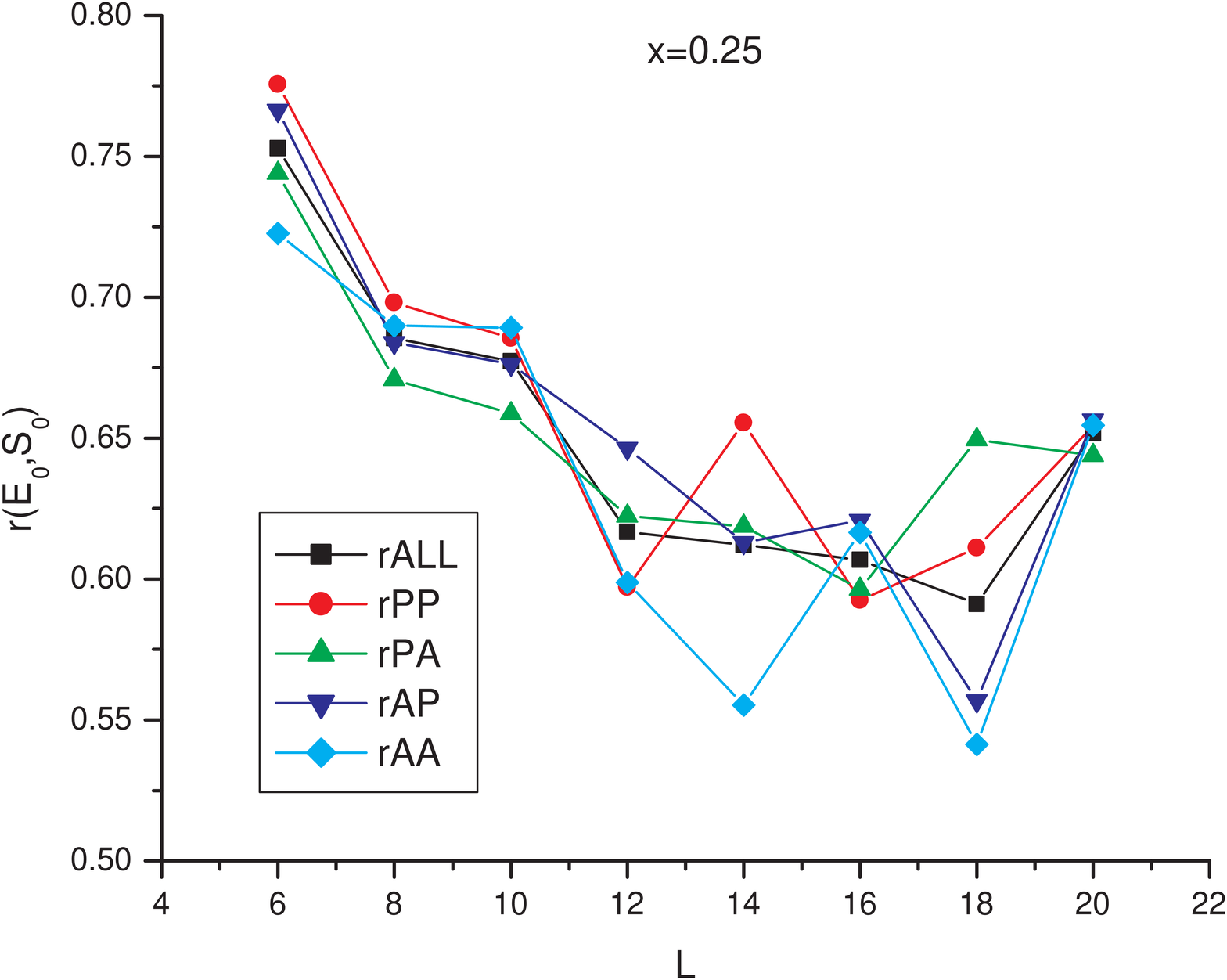}
\caption{\label{Fig.6}(color online) Covariance of ground state
energy and ground state entropy, $r(E_0,S_0)$ vs. $L$ for $x = 0.25$.}
\end{figure}

\begin{figure}
\includegraphics[width=3in]{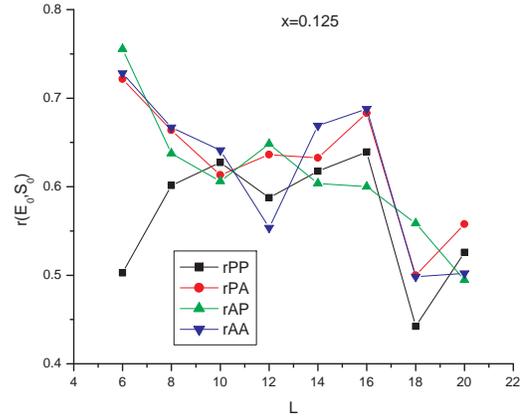}
\caption{\label{Fig.7}(color online) Covariance of ground state
energy and ground state entropy, $r(E_0,S_0)$ vs. $L$ for $x = 0.125$.}
\end{figure}

In Figures 5, 6 and 7, we give the value of $r$, the
normalized covariance defined by
\begin{equation}
  r ( E_0 , S_0 ) = {{ \langle E_0 S_0 \rangle - \langle E_0 \rangle
  \langle S_0 \rangle } \over { \sigma ( E_0 ) \sigma ( S_0 )}}  \, ,
\end{equation}
for each fit.  The angle brackets indicate an average over the
random bond distribution for some fixed value of $x$.  The
standard deviation, $\sigma$, is defined, as usual, as
\begin{equation}
  \sigma (X) = \sqrt{ \langle X^2 \rangle - {\langle X \rangle}^2 }  \, .
\end{equation}
The numbers in parentheses and the error bars shown in the figures
represent a one standard deviation statistical error, as
calculated by the Origin 6.0 Professional\cite{Ori} least-squares
fitting routine.  One expects that, in addition, there may be
systematic errors arising from non-ideal behavior of random number
generators and nonlinear correlations.  It is often difficult to
obtain meaningful estimates of systematic errors.

For small $L$ there is a strong correlation between $E_0$ and
$S_0$. As $L$ increases, $\sigma (E_0)$ increases linearly with
$L$ but $\sigma (S­_0)$ increases faster than linearly over this
range of $L$.  Thus the correlation gets weaker as $L$ increases.
This is reflected in the tendency for $r$ to decrease as $L$
increases. From our data, it is not clear whether or not $r$ goes
to zero as $L$ goes to infinity.  It is generally believed that
the model is not self-averaging at $T = 0$, so it would be natural
for $r$ to remain finite as $L$ increases.

The value of $r$ depends on the choice of ensemble.  For the
ensemble in which we fix both the number of negative bonds and the
number of frustrated plaquettes for each value of $L$, one would
find higher values of $r$ than what we find here.  Crudely, one
would expect the values of $r$ in the more tightly specified
ensemble to be higher by about a factor of $1 / .8$, the inverse
of the $r$-factor for the correlation between $E_0$ and the number
of frustrated plaquettes.

For $x = 0.25$ and $x = 0.5$, as $L$ increases the slope of the
regression line through the data given by the least-squares fit
appears to rapidly approach a limit of about $m \approx 0.36$.
This number is slightly greater than $ln(2)/2 = 0.34657...$.
Naively, this means is that, on the average, the GS degeneracy
increases by about a factor of two for each additional broken
bond, since each broken bond increases the energy by two units.
But $m$ is not actually a physical observable, because it depends
on our choice of ensemble.  We will say more about what the
actual physical quantity is later.

\begin{figure}
\includegraphics[width=3in]{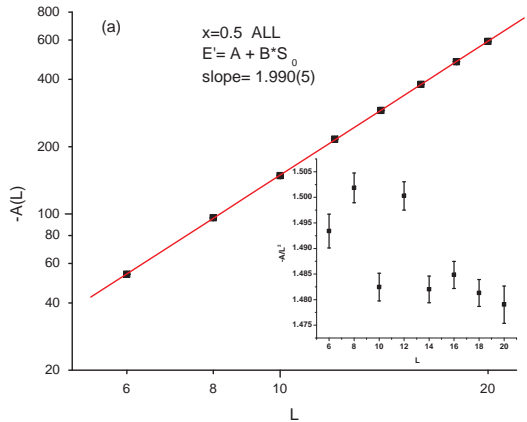}\quad
\includegraphics[width=3in]{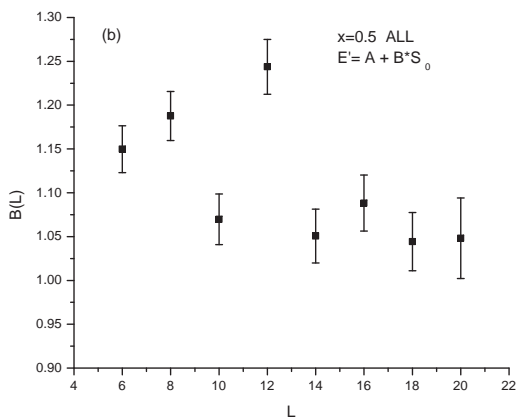}
\caption{\label{Fig.8}(color online) Results of least-squares fit
analysis parameterized by Eqn.~(6) of the scatter plot of
correlations between $E_0$ and $S_0$ for $x = 0.5$: (a)$-A$ vs.
$L$, log-log plot; inset: $-A / L^2$ vs. $L$; (b)$B$ vs. $L$.}
\end{figure}

\begin{figure}
\includegraphics[width=3in]{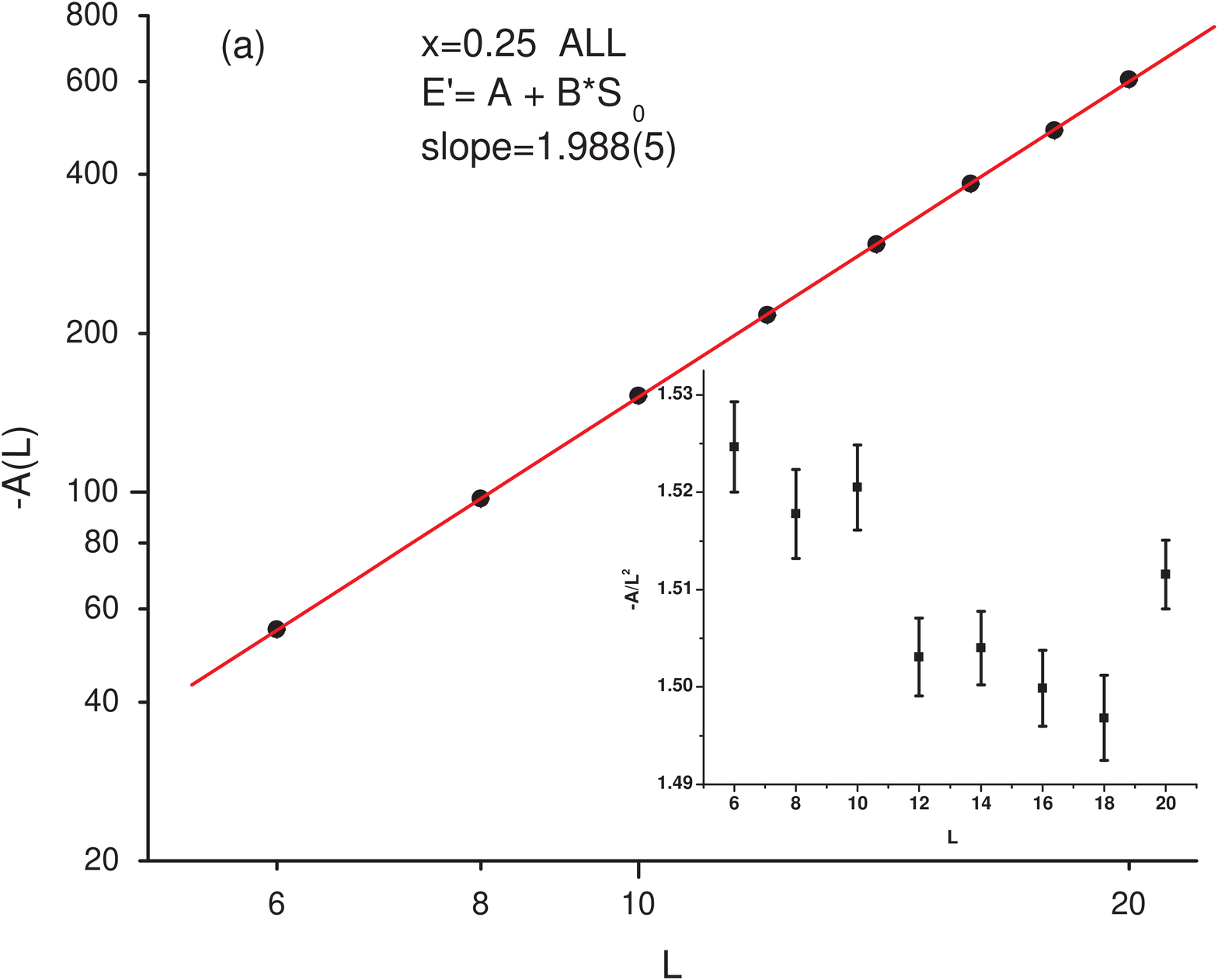}\quad
\includegraphics[width=3in]{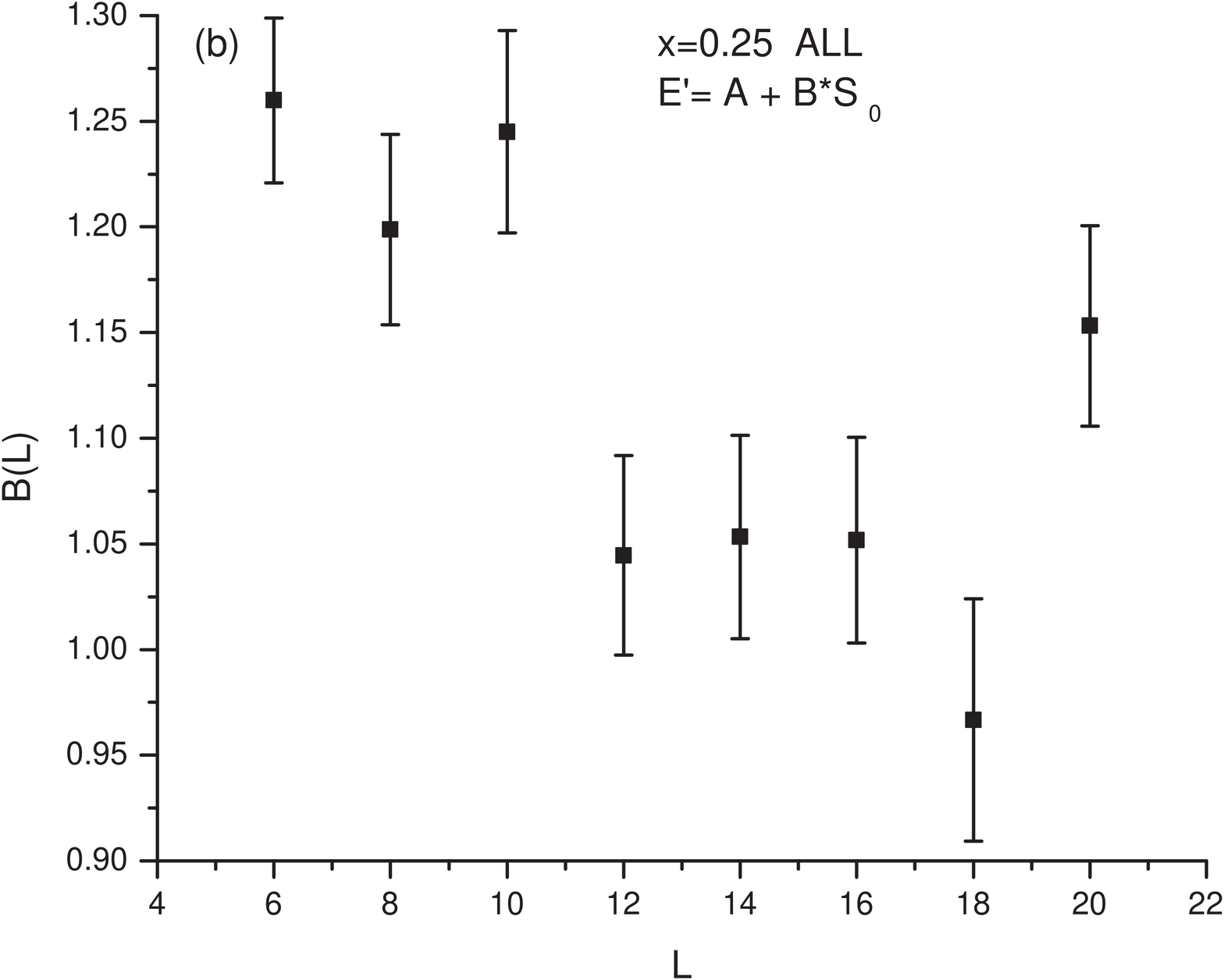}
\caption{\label{Fig.9}(color online) Results of least-squares fit
analysis parameterized by Eqn.~(6) of the scatter plot of
correlations between $E_0$ and $S_0$ for $x = 0.25$: (a)$-A$ vs.
$L$, log-log plot; inset: $-A / L^2$ vs. $L$; (b)$B$ vs. $L$.}
\end{figure}

\begin{figure}
\includegraphics[width=3in]{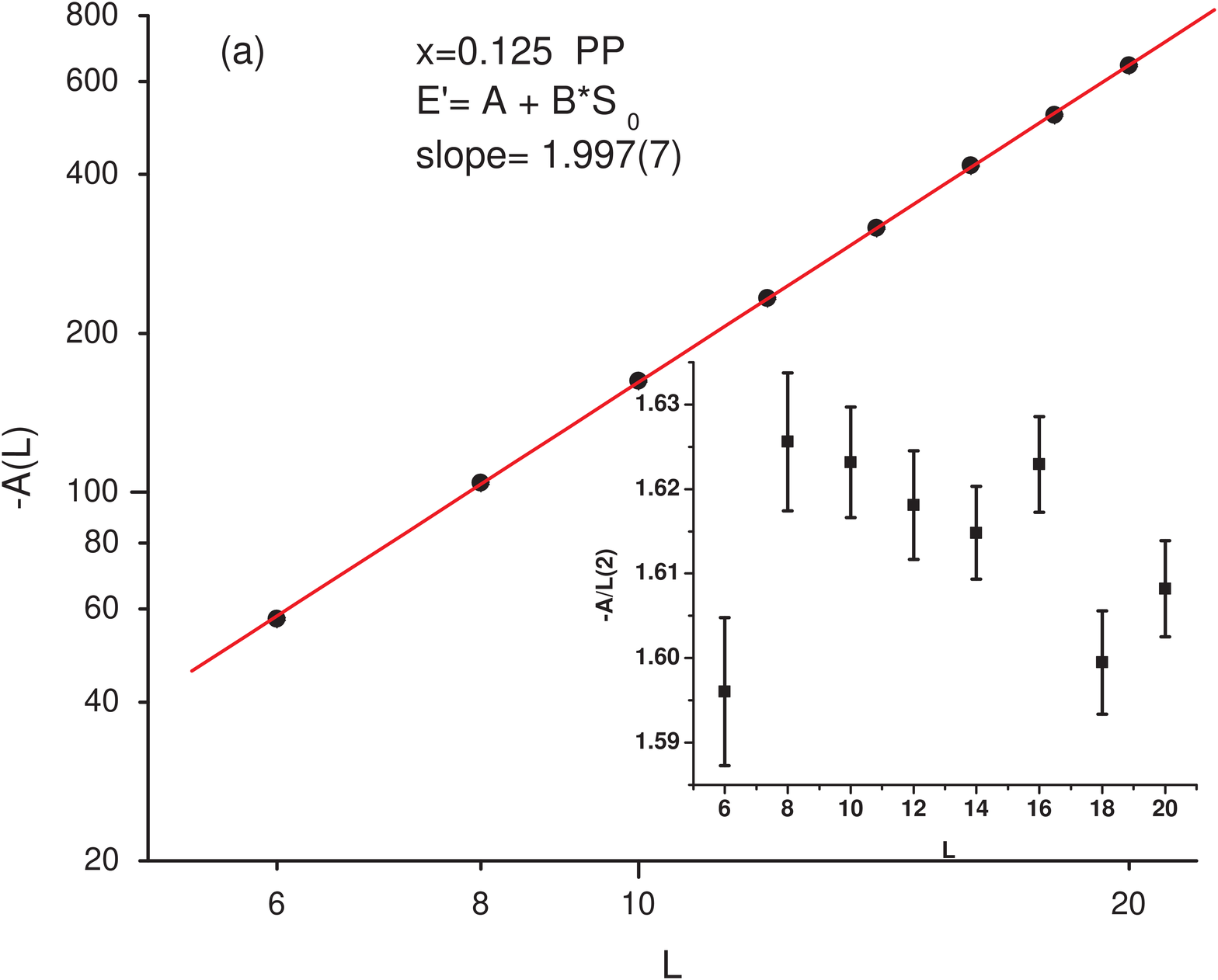}\quad
\includegraphics[width=3in]{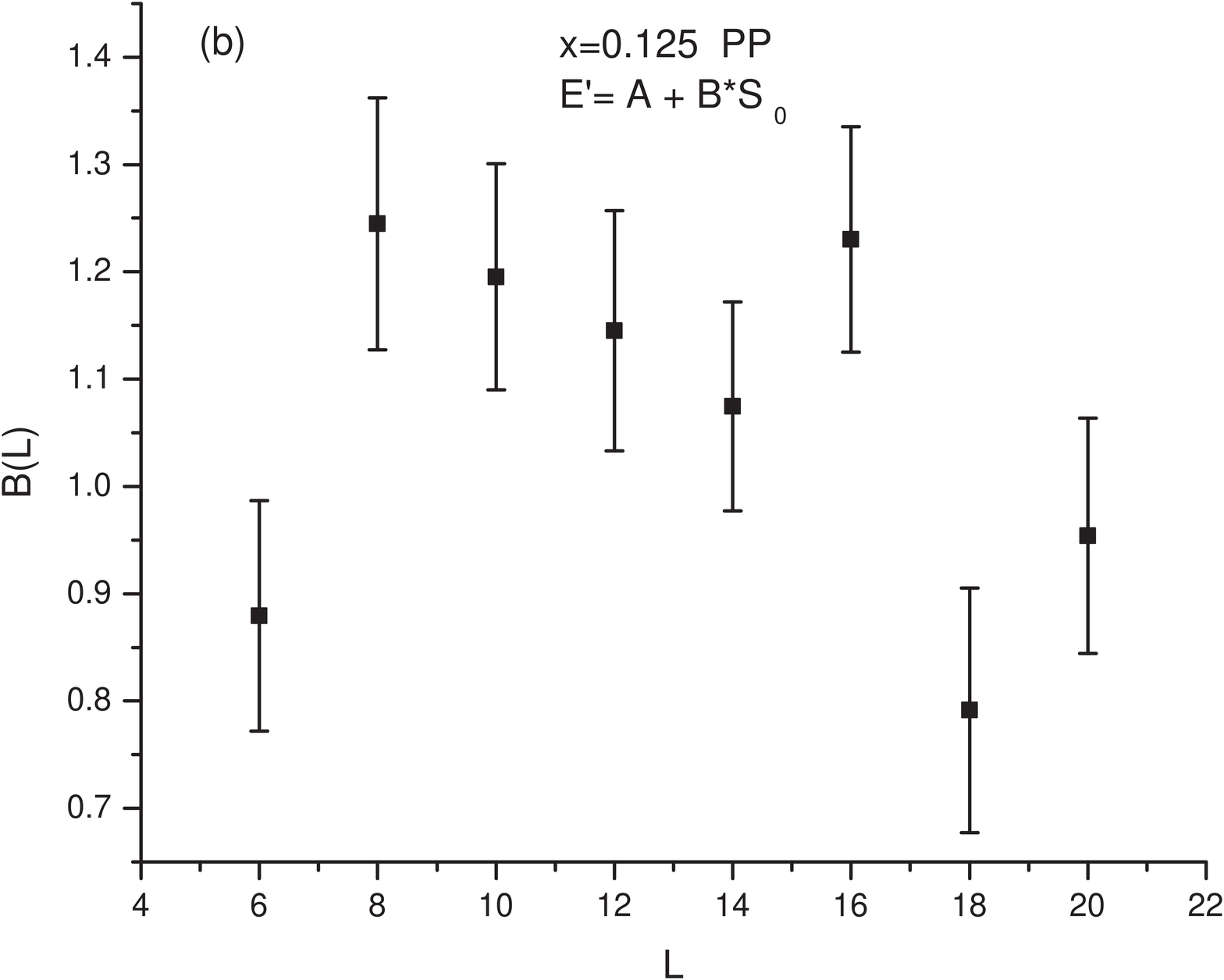}
\caption{\label{Fig.10}(color online) Results of least-squares fit
analysis parameterized by Eqn.~(6) of the scatter plot of
correlations between $E_0$ and $S_0$ for $x = 0.125$: (a)$-A$ vs.
$L$, log-log plot; inset: $-A / L^2$ vs. $L$; (b)$B$ vs. $L$.}
\end{figure}

The reader should note that the probability density in the
energy-entropy plane shown in Fig.~1(c) is clearly different from
a two-dimensional Gaussian distribution, since, with the boundary
conditions we are using, $E_0$ can only have values which are
multiples of four units.  The previous results in Figure~6 of
Landry and Coppersmith,\cite{LC01} using a much larger number of
samples, show that the one-dimensional probability distribution
for $S_0$ at the same values of $x$ and $L$, which is the
projection of the joint distribution onto the entropy axis, can be
fit by a Gaussian distribution.

Since it is generally believed that $T = 0$ is a critical point
for the model, perhaps what one needs to explain is why the
one-dimensional distribution is apparently Gaussian!  This result
becomes less surprising, however, once one realizes that the
strong correlations between $E_0$ and the number of frustrated
plaquettes will make it very difficult to see any non-Gaussian
behavior in this one-dimensional distribution, unless one holds
the number of frustrated plaquettes fixed.  In our ensemble, the
number of frustrated plaquettes always has a Gaussian
distribution.

For $x = 0.125$, where the ferromagnetic correlations are
substantial\cite{AH04} for small $L$, the strength of the
energy-entropy correlation is somewhat reduced for the case of
periodic boundary conditions in both directions.  This effect is
probably a result of the fact that for small $L$ at $x = 0.125$,
the behavior is essentially dominated by short-range ferromagnetic
correlations.

It is equally valid to do the least-squares fit using $S_0$ as the
independent variable. The results of fits of this type for $x =
0.5$, 0.25 and 0.125, using the same data as before, are shown in
Figures 8, 9 and 10, respectively.  The values of $r$ are not
shown again, since they are unchanged from the earlier case. For
each value of $x$ and each $L$, we now show the slope $B$ of the
least-squares fit, and the offset $A$ of the energy, defined by
\begin{equation}
  E' (S_0) = A + B*S_0  \, .
\end{equation}
Now the notation $E'$ indicates that energy is the dependent
variable in the least-squares fit.  Since $A$ is negative, we use
$-A$ in the log-log plot.

From the log-log plots in Figs.~2(a) and 3(a), we find that $b$ is
proportional to $L^{1.78(2)}$ and $L^{1.81(4)}$, respectively,
while from Figs.~8(a) and 9(a), the scaling of $-A$ is essentially
indistinguishable from $L^2$.  A closer inspection of Fig.~3(a),
however, reveals that there is a clear curvature of the data on
the log-log plot.  We can see in the inset to Fig.~3(a) that, for
$x = 0.25$, $b$ becomes essentially proportional to $L^2$ for $L
\ge 12$.

There is no known reason for a qualitative difference in the
scaling behavior between $x = 0.5$ and $x = 0.25$.  Therefore, we
anticipate that the entropy scaling will also become proportional
to $L^2$ for some larger $L$ in the $x = 0.5$ case.  Given the
results in the literature, we expect that this will happen before
$L$ reaches 30.

As we have demonstrated in Fig.~1(c) and 1(d), the regression line
which is obtained when one uses $S_0$ as the independent variable
is not the same one which is found by using $E_0$ as the
independent variable. The slope parameters of these two regression
lines are related to each other\cite{Tu51} as
\begin{equation}
  m B = r^2  \, .
\end{equation}
Since $r^2$ is, in general, a number between 0 and 1, we are faced
with the problem of deciding what the true best line through the
data is. Without some additional information, there is no unique
prescription for solving this problem.\cite{Ma59}

We can write down an equation for the joint probability
distribution which builds in the fact that the allowed values of
$E_0$ are quantized:
\begin{widetext}
\begin{equation}
  P_L ( E_0 , S_0 )  = C_L f {\big (}{{E_0 - \langle E_0(L) \rangle}
  \over
     {\sigma ( E_0(L) )}}, {{S_0 - \langle S_0(L) \rangle} \over
     {\sigma ( S_0(L) )}}{\big )}
     [ \sum_{n = - \infty}^\infty \delta ( E_0 - 4 n ) ] \, .
\end{equation}
\end{widetext}
The dependence on $x$ is not shown explicitly, and $C_L$ is the
normalization constant.  Aside from small corrections to scaling
which can be ignored for large $L$, we expect that we can assume
\begin{equation}
\langle E_0(L) \rangle = E_\infty L^2 \, ,
\end{equation}
and
\begin{equation}
\langle S_0(L) \rangle = S_\infty L^2 \, .
\end{equation}
It should also be safe\cite{SK94,LC01} to assume that $\sigma(
E_0(L) ) =  \bar{\sigma}_E L$.  The scaling of $\sigma( S_0(L) )$
with $L$ appears to be nontrivial,\cite{SK94,LC01} but we
certainly expect that it will diverge as $L$ goes to infinity.
Therefore, for large $L$ it should be an adequate approximation to
replace the sum over the $\delta$-functions by a uniform
background.

If the envelope function $f$ of the probability distribution in
the energy-entropy plane was a two-dimensional Gaussian, then it
would have the normal form
%\begin{widetext}
\begin{equation}
  f_G ( X , Y ) = {1 \over {2 \pi \sqrt{1 - r^2}}} \exp {\big (}
   - {{ X^2 -2 r X Y + Y^2 } \over {2 ( 1 - r^2 )}}{\big )} \, ,
\end{equation}
%\end{widetext}
where the arguments $X$ and $Y$ have probability distributions
with zero mean and unit standard deviation.  Given this
assumption, which is unproven, we should treat $E_0$ and $S_0$ on
an equal basis. Then it would be correct to set the best
regression line through the joint probability distribution to be
equal to
\begin{equation}
  S_0 - S_\infty = (m/r)*( E_0 - E_\infty ) \, ,
\end{equation}
or, equivalently,
\begin{equation}
  E_0 - E_\infty = (B/r)*( S_0 - S_\infty ) \, .
\end{equation}
As we have repeatedly reminded the reader, $T = 0$ is believed to
be a critical point for this model, so the assumption of Gaussian
fluctuations can be justified only as an approximation. We do not
really know what actual form of $f ( X , Y )$ should be used, and
therefore\cite{Tu51,Ma59} we do not know what the slope of the
best regression line should be.  However, it seems certain that
corrections to the Gaussian approximation are invisible at our
current level of statistical uncertainty.

\section{DISCUSSION}

One might think that the behavior of the $A$ and $B$ parameters
which we find by treating the entropy as the independent variable
in the least-squares fit are quite simple.  It is important to
remember, however, that the slope of the best line through the
data is {\it not} $B$.  Within the Gaussian approximation, as we
have remarked above, the slope of the best line is $B / r$.  And
thus if the slope of the best line through the data is to have
some finite slope in the large $L$ limit, it appears necessary to
have a finite limit for $r$.

In Figures 11, 12 and 13, we show the values of $B / r$ vs. $L$.
The value of $B / r$ appears to be approximately independent of
$L$ for $x$ = 0.125 and 0.25, because the $L$ dependences of $B$
and $r$ cancel each other, although the statistical uncertainties
are too large for precise statements to be made.  For $x$ = 0.5,
however, $B / r$ may be increasing with $L$ within our range of
$L$.

Dimensionally, the slope $B / r$ defined in Eqn.~(13) has units of
temperature. It is tempting to argue that $B / r$ has some
relation to a fictive glass temperature for the crossover between
high and low temperature dynamical behavior.  Thus, a naive
prediction for $B / r$ would be that it should be proportional to
the mean-field energy scale, which is
\begin{equation}
  E_{mf} = 2 \sqrt{x (1 - x)}  \, ,
\end{equation}
where the factor of 2 comes from the square root of the number of
neighbors on the lattice, and the factor of $x (1 - x)$ is the
second moment of $P ( J_{ij})$.  However, no such dependence on
$x$ is seen in our data.  The value of $B / r$ actually seems to
be decreasing slowly as $x$ increases from 0.125 to 0.5.  For $x =
0.5$, if we average the data for all $L$, we find
\begin{equation}
  B / r = 1.66 \pm 0.03  \, .
\end{equation}
This number probably underestimates the result for large $L$
slightly, due to the apparent tendency for $B / r$ to increase
with $L$.  The quoted statistical error does not include any
allowance for this effect.  Using Eqn.~(15), however, we find that,
on the average, each additional broken bond in the ground state
increases the GS degeneracy by a factor of 3.34(7).  Our
uncertainties for the smaller values of $x$ are larger, but this
is primarily because we have smaller numbers of samples for these
cases.

\begin{figure}
\includegraphics[width=3in]{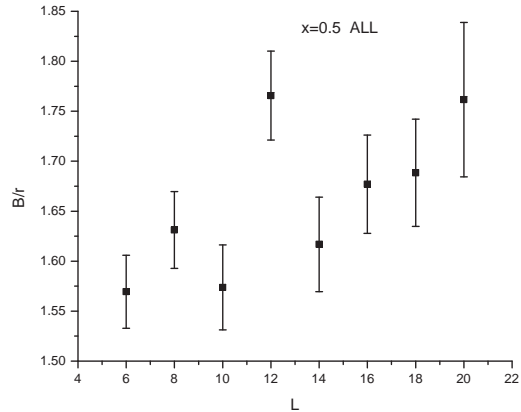}
\caption{\label{Fig.11} $B / r$ vs. $L$ for $x = 0.5$.}
\end{figure}

\begin{figure}
\includegraphics[width=3in]{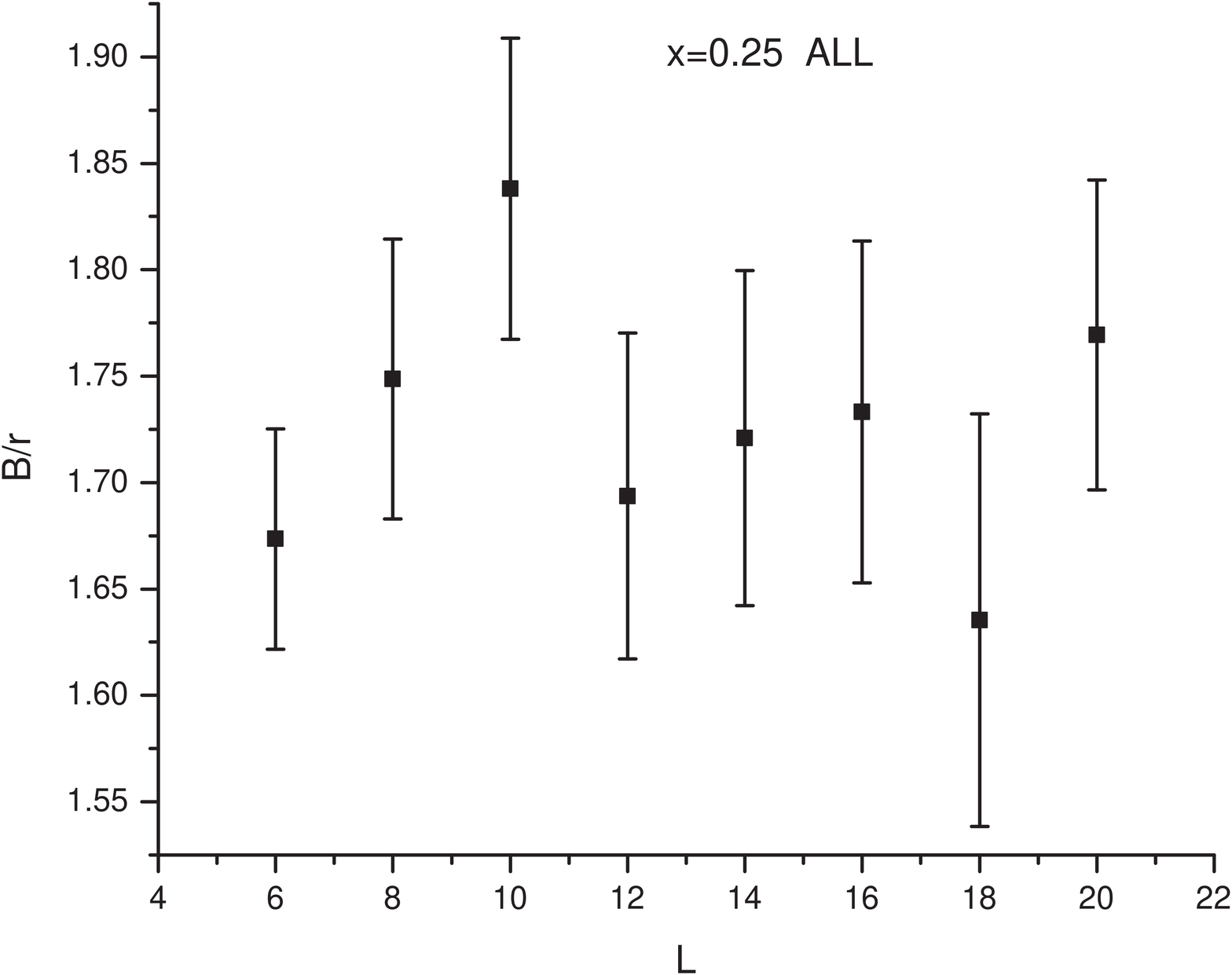}
\caption{\label{Fig.12} $B / r$ vs. $L$ for $x = 0.25$.}
\end{figure}

\begin{figure}
\includegraphics[width=3in]{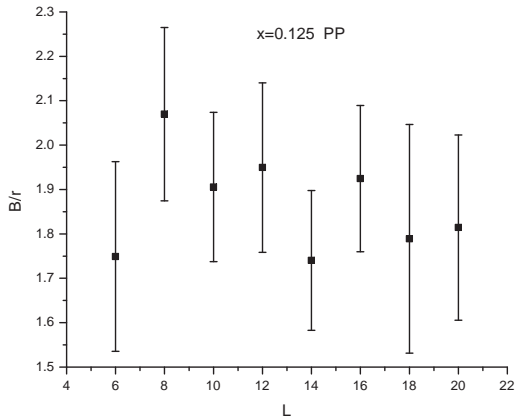}
\caption{\label{Fig.13} $B / r$ vs. $L$ for $x = 0.125$.}
\end{figure}

For any GS on a square lattice, each spin which has two neighbor
spins that are oriented along the direction favored by the bond
between them, with the other two neighbor spins pointed opposite
to the direction favored by the bond ({\it i.e.} these bonds are
broken), can flip with no energy cost.  Each of these free spins
contributes a factor of two to the degeneracy of the GS.  It is
thus expected that increasing the number of broken bonds in the GS
would also increase the number of such free spins. It is not
reasonable, however, that increasing the number of broken bonds by
one would increase the number of free spins by nearly two, on the
average.  Therefore, the fact that we have found the average
increase in the GS degeneracy for each additional broken bond to
be a factor of about 10/3 indicates that this effect cannot be
explained by fluctuations in the number of zero-energy single-spin
flips.  There must be a substantial contribution from large-scale
rearrangements of the GS structure.

An exact calculation of the energy-entropy correlation for $L =
50$, or possibly $L = 60$, could be performed using the method of
Galluccio, Loebl and Vondr\'{a}k.\cite{GLV00} It should be noted,
however, that it is not really necessary to calculate $S_0$
exactly.  It would be more than sufficient to have an approximate
calculation of $S_0$ which was accurate to one part in $10^4$.
That does not seem impossible, and it might allow calculations for
even larger $L$.

An explicit calculation of the low temperature specific heat for
$x = 0.5$ by Lukic {\it et al.}\cite{LGMMR04} gives a result
proportional to $\exp( -2 / T )$ when $L > 30$.  It is natural
that crossover behavior should be seen in both $S_0$ and the low
$T$ specific heat, with the same crossover length.  The low $T$
limit of the specific heat in this model is determined by the
degeneracy of the states at energy 4 above the GS.  It is surely
not surprising that the degeneracies of the ground states and the
first excited states would be controlled by the same crossover
length.  To verify that this is occurring, the calculation of
Lukic {\it et al.} could be repeated for $x = 0.25$. We expect
that a crossover length of $L \approx 12$ will be found for the
specific heat in that case.

Katzgraber and Lee\cite{KL05} have calculated the $T$ dependence
of the correlation length in this model, and found that it behaves
as $\exp ( 2 / T )$.  (Recall that $J = 1$.)  They use this result
to argue that the specific heat at low $T$ should be proportional
to $\exp ( -4 / T )$, as one might naively expect for a model with
an energy gap of 4.  However, a more detailed analysis\cite{KLC06}
has found that their data for the specific heat agree with the
conclusions of Lukic {\it et al.}\cite{LGMMR04}  Another recent
study by Wang,\cite{Wan05} using a new algorithm, also finds that
the low $T$ specific heat is proportional to $\exp( -2 / T )$.

It would also be interesting to repeat these calculations on a
hexagonal lattice, where the allowed energy states are multiples
of two units, because the number of bonds for each site is odd.
Thus on this lattice the smallest zero-energy excitation of the
$\pm J$ model involves two neighboring spins.  We would expect
that the low temperature specific heat is proportional to $\exp
( -2 / T )$ for the whole range of $L$ in that case.

The analogy to an Ising chain which is made by Wang and
Swendsen\cite{WS88} to argue for a specific heat which is
proportional to $\exp ( -2 / T )$ has nothing to do with random
bonds.  We know, however, that in 2D a fully frustrated Ising
system does not display this behavior.\cite{SK94}  In addition, by
studying triangular lattices, Poulter and Blackman\cite{PB01}
found that adding a small concentration of unfrustrated plaquettes
to a fully frustrated system does not produce spin-glass behavior.

It was recently shown by Amoruso, Marinari, Martin and
Pagnani\cite{AMMP03} that the behavior of domain wall energies for
the 2D Ising spin glass is fundamentally different in those cases,
such as the $\pm J$ model, where the energies are quantized.  In a
very interesting paper, Wang, Harrington and Preskill\cite{WHP03}
have argued that the presence of an energy gap allows the
existence of topological long-range order.  A less specific
suggestion of topological long-range order in 2D random-bond Ising
models was made earlier by Merz and Chalker.\cite{MC02}  Numerical
results for the properties of domain walls, which will be
presented elsewhere,\cite{Fis05,FH06} are consistent with this
proposal.

What we want to do is to explain the low temperature behavior of
the specific heat in terms of topological excitations.  We know
that in the spin-glass region of the phase diagram it is difficult
to find a way to overturn a finite fraction of a large sample at
zero energy cost.  In contrast, at and near the fully frustrated
system, where the $\exp ( -2 / T )$ behavior of the specific heat
does not occur, it becomes easy to find ways of overturning a
finite fraction of the spins at zero energy cost.  This explains
the difference between the spin-glass region and the fully
frustrated region.

What is possible in the spin-glass region is finding many ways of
overturning a finite fraction of the large sample at a cost of
only 4.\cite{Fis05}  This is precisely the generalization of the
Wang-Swendsen mechanism\cite{WS88} to a random 2D system.  And
because we can do this, the behavior of the Parisi overlap
function\cite{Par83} will be nontrivial.  Of course, since $T_c$
is zero, the overlap function collapses to zero as $L$ increases.
But if we scale out this simple collapse with $L$, the existence
of the large-scale finite-energy excitations may be observed in
the overlap function.  Since we are trying to observe effects
caused by states of energy $4 J$, we cannot merely do a naive $T =
0$ calculation of the overlap function.  However, by manipulating
the double limit $L \rightarrow \infty$ and $T \rightarrow 0$, it
may be possible to see the effect.

Although the actual implementation would be very challenging, one
can imagine studying the energy-entropy correlation at and below
$T_c$ in a three-dimensional Ising spin glass, using
thermal-average values for the energy and entropy. In that case,
where the spin-glass phase and the failure of self-averaging of
the Parisi overlap function are believed to occur at finite
$T$,\cite{PSR03} one can use a general probability distribution
for $P ( J_{ij})$ and still have a positive entropy, in contrast
to the 2D situation.

The author's expectation is that the anomalous scaling which we
find for the $\pm J$ model at $T = 0$ in 2D  for small $L$ will
occur for all types of bond distributions in 3D, where the
spin-glass transition is at $T > 0$.  It might also happen that
the crossover length becomes infinite in 3D, but it seems more
likely that the crossover length is only infinite in four or more
dimensions.

\section{SUMMARY}

We have found that for $L \times L$ square lattices with $L \le
20$ the 2D Ising spin glass with +1 and -1 bonds has a very strong
correlation between $E_0$ and the number of frustrated plaquettes,
and, what is more surprising, a strong correlation between $E_0$
and $S_0$. On the average, each additional broken bond in the GS
of a particular sample of random bonds increases the GS degeneracy
by a factor of about 10/3. This number is too large to be
explained by fluctuations in the number of free spins, which
implies that there is a substantial contribution due to
large-scale rearrangements of the GS structure. Over this range of
$L$, the characteristic GS entropy scales as $L^{1.78(2)}$ for $x
= 0.5$, while the characteristic GS energy scales as $L^2$, as
expected. For $x = 0.25$, however, a crossover is seen to normal
scaling behavior of $S_0$ near $L = 12$. We believe that a similar
crossover will occur for $x = 0.5$ at $L \approx 25$, and that
this crossover is connected to the anomalous behavior of the low
temperature specific heat.  We explain why the Wang-Swendsen
mechanism for a low $T$ specific heat proportional to $\exp( -2 /
T )$ should apply in the spin-glass regime, but not in the fully
frustrated regime.

\begin{acknowledgments}
The author thanks S. N. Coppersmith for generously providing all of
the raw data analyzed in this work, and for a careful reading of an
early version of the manuscript.  The computer program used by Prof.
Coppersmith to generate the data was written by J. W. Landry, and is
an adaptation of the code written by L. Saul.  The author is grateful
to S. L. Sondhi, F. D. M. Haldane, A. K. Hartmann and H. G.
Katzgraber for helpful discussions, and to Princeton University for
providing use of facilities.

\end{acknowledgments}


%\newpage

\end{document}